\documentstyle[pre,aps]{revtex}

\begin{document}

\draft

\title{Unified Approach to Crossover Phenomena} 

\author{S. Gluzman$^1$ and V. I. Yukalov$^{2,3}$\footnote{Corresponding
author}}

\address{$^1$International Centre of Condensed Matter Physics\\
University of Brasilia, CP 04513, Brasilia, DF 70919-970, Brazil\\
$^2$Centre for Interdisciplinary Studies in Chemical Physics\\
University of Western Ontario, London, Ontario N6A 3K7, Canada \\
$^3$ Bogolubov Laboratory of Theoretical Physics\\
Joint Institute for Nuclear Research, Dubna 141980, Russia}

\maketitle

\begin{abstract}

A general analytical method is developed for describing crossover
phenomena of arbitrary nature. The method is based on the algebraic
self--similar renormalization of asymptotic series, with control 
functions defined by crossover conditions. The method can be employed 
for such difficult problems for which only a few terms of asymptotic
expansions are available, and no other techniques are applicable. As 
an illustration, analytical solutions for several important physical
problems are presented.
\end{abstract}

\vspace{1cm}

\pacs{02.30.Lt, 05.20--y, 11.10.Hi}

\section{Introduction}

Crossover phenomena are ubiquitous in nature. Probably, they are much more
common than phase transitions. When speaking about crossover phenomena,
one usually keeps in mind the following picture: A function $f(x)$,
describing a physical quantity, is continuous in an interval $x_1\leq
x\leq x_2$, but the behaviour of $f(x)$ in the vicinity of the boundaries
of this interval is {\it qualitatively different} near $x_1$ as compared
to $x_2$. The qualitative change of the behaviour of this function, as $x$
moves from one side to another side is commonly understood as a crossover.

It is possible to quote hundreds of examples of different crossovers. For
instance, many physical quantities qualitatively change their behaviour
when passing from the weak--coupling to strong--coupling limit [1]. This
concerns, e.g., the majority of problems having to do with the behaviour
of energies as functions of a coupling parameter in statistical physics,
quantum mechanics, and field theory. Let us mention in this respect the
dependence of the spectra of Schr\"odinger operators on the anharmonicity
parameter for variegated anharmonic models. The energy spectrum of such
models is qualitatively different in the weak--coupling (weak anharmonicity) 
as compared to the strong--coupling (strong anharmonicity) limits.

A famous example of a crossover phenomenon is the Kondo effect [2] when
the behaviour of a system changes qualitatively at varying temperature.
Although this transformation goes smoothly, with no discontinuities in
thermodynamic characteristics, but the change of properties is so
noticeable that one can ascribe a particular point, called the Kondo
temperature, to a region dividing qualitatively different regimes of low
and high temperatures.

Another renowned example of a crossover is the Fr\"ohlich polaron problem
[3]. Polaron characteristics, such as its energy or effective mass, change
qualitatively when varying the coupling parameter describing electron--phonon
interactions. This change happens so explicitly that for about two decades
there were many speculations suggesting that there exists a phase
transition at a particular value of the coupling parameter. However,
modern highly accurate Monte Carlo calculations [4] confirm the initial
Feynman picture [3] proving that we meet here not a phase transition but a
classical crossover.

In the examples mentioned above, of simple anharmonic models, the Kondo
effect, and of the Fr\"ohlich polaron problem, the crossover, when varying
a coupling parameter or temperature, is monotonic. However, there are
cases when crossover is not monotonic. This concerns, for instance,
one--dimensional antiferromagnet whose characteristics are considered as
functions of spin. Then the Haldane gap [5], as spin changes from small
to large values exhibits a very nonmonotonic behaviour becoming zero at
each half--odd--integer spin.

A nonmonotonic behaviour can often be met in the dependence of spectra 
of collective excitations on wave vector. Then the crossover from the
region of small wave numbers, corresponding to the long--wavelength
acoustic regime, to the region of large wave vectors, corresponding to a
single--particle regime, can go through a nontrivial intermediate region
displaying maxima and minima, associated with maxons and rotons [6-8].

We could adduce a number of other examples of crossover phenomena related
to interesting physical problems. Let us just mention deconfinement in
nuclear matter, which is rather a crossover phenomenon than a phase
transition (see discussion in review [9]). However, we think, it is already 
clear that crossover phenomena are widespread in nature and that it is
important to know how to describe them.

The description of crossover phenomena occurring in realistic statistical
systems is usually very complicated. This is because one needs to find
physical characteristics for a wide range of parameters, which is far from
being trivial for  complex systems. Say, we have to find a functions
$f(x)$ on the whole semi--axis $0\leq x < \infty$. The variable $x$
may represent, e.g., a coupling parameter, temperature, or wave vector.
Quite often, one can define, more or less easily, the asymptotic behaviour
of $f(x)$ near the boundaries of the interval $[0,\infty)$, that is, when
$x\rightarrow 0$ and $x\rightarrow\infty$. Such asymptotic expressions may
correspond to the weak--coupling and strong coupling limits, to the
low--temperature and high--temperature expansions, to the long--wavelength
and short--wavelength approximations, and so on. But in the intermediate
region, where there are no small parameters, one cannot invoke perturbative 
techniques. It would be nice to possess  a method allowing to construct
interpolation formulas only from the knowledge of asymptotic expansions
near boundaries.

There exist several summation techniques, such as Pad\'e approximation,
Borel summation, conformal mapping, and so on, that permit to ascribe
effective sums to asymptotic series [1,10]. But all these techniques are
not applicable in principle for the complex problems we are interested in
here. This is because of the following main reasons: First, all these
summation techniques, to be applicable, require the knowledge of tens of
terms in an asymptotic series. Such a luxurious information is usually not
available for nontrivial systems, for which standardly one is able to
derive just a few perturbative terms. Then all mentioned summation
techniques are useless. Second, the latter are just {\it summation} methods, 
while we here are concerned about an {\it interpolation} problem.
Summation and interpolation are far from being the same!

The most known interpolation method is the {\it two--point} Pad\'e
approximation [11], that should not be confused with the standard Pad\'e
approximation [10]. However, the former method, being a derivative of the
latter, shares all its deficiences. Among the most important shortcomings
of Pad\'e approximants, we may mention the following: the necessity of
having many perturbative terms, the appearance of unphysical poles, the
ability of dealing only with the so--called compatible variables, the
possibility of describing only those functions that have at infinity a
power--law behaviour with rational powers, and the impossibility to
correctly treat nonmonotonic crossover. These difficulties are well known
and repeatedly discussed in literature [10-18]. In addition, we remind
that Pad\'e approximation is rather a numerical method.

In the present paper, we advance an {\it analytical} approach for 
treating interpolation problems of arbitrary nature. This approach is free
of the shortcomings typical of the two--point Pad\'e approximation. And
what also makes this approach more general than any other know methods is
the possibility of using it for those difficult cases when just a few
asymptotic terms are available and no other method is applicable. We
illustrate the approach by applying it to several difficult problems with
monotonic as well as with nonmonotonic crossover. We would like to stress
that the physical problems we consider not only illustrate the wide
applicability of the suggested approach but are also of interest as such.
Therefore, the interpolation formulas we derive present analytical
solutions for important physical problems.

\section{General Approach}

The interpolation approach we advance here is based on the ideas of our
previous papers. However, since we do not assume that a reader in the
common audience is already well acquainted with these ideas, we provide
here a clearly understandable description of the method in general,
complementing it by those particulars that are necessary for adjusting 
it to the interpolation problem

Assume that we are looking for a physical characteristic presented by a
function $f(x)$, in which the variable $x$ changes in the interval
$[0,\infty)$. The standard situation is when the physical problem under
investigation is so complicated that it is difficult, or even impossible,
to find a reasonable approximation for the sought function in the whole
given interval. However, it is often feasible to get an asymptotic
expansion for small variables
\begin{equation}
\label{1}
f(x)\simeq p_k\ (x), \qquad ( x\rightarrow 0) ,
\end{equation}
where $k=0,1,2,\ldots$, employing a kind of perturbation theory. Also, it
is often possible to find an asymptotic behaviour of the function at
large variables, say,
\begin{equation}
\label{2}
f(x) \simeq f_{as}(x) \qquad (x\rightarrow \infty) .
\end{equation}
Then the interpolation problem consists in answering the question: What
can be said about the behaviour of the function in the whole interval
$[0,\infty)$ being based on the asymptotic information in (1) and (2)?
Usually, not much, since the asymptotic expressions (1) and (2), being
derived in two opposite limits, have nothing common with each other. In
addition, perturbative approximations, such as (1), usually result in
divergent series. When one is lucky enough, dealing with a more or less
easy case, so that tens of perturbative terms in (1) could be calculated,
then one could invoke some known summation technique in order to ascribe
an effective sum to a divergent series. However, even in such a lucky
case, the found effective sum may have, and usually has nothing to do with
the limit (2) from another side of the axis $[0,\infty)$. For example, the
standard case is when the limit (2) corresponds to an exponential
behaviour. If one uses Pad\'e approximants or any other techniques based
on them, for instance, Pad\'e--Borel summation [1], one comes to an
effective sum in the form of rational fractions, which cannot be matched
with an exponential. In the less lucky but more realistic case, when only
a few perturbative terms are known, all these summation techniques become
in principle useless. How could we proceed in such difficult cases in
order to find an interpolation formula connecting (1) and (2)?

The first thing we need to do is to understand how to extract a useful
information from a divergent sequence $\{ p_k(x)\}$ when only a few
initial terms of it are available. It would be nice to reconstruct the
sequence $\{ p_k(x)\}$ in such a way that to improve its convergence
properties. Having only a few terms, we cannot resort to the standard
summation techniques. Nevertheless, a reconstruction is possible with the
help of {\it control functions} [19,20]. Let us denote the procedure of
introducing control functions as
\begin{equation}
\label{3}
C_s\{ p_k(x) \} = P_k(x,s) , 
\end{equation}
where $s=s_k(x)$ is a set of functions such that the sequence 
$\{ P_k(x,s_k)\}$ has better convergence properties than $\{ p_k(x)\}$.
The name "control functions" reflects their role in controlling
convergence. The introduction of such functions can be done in several
ways. Generally, any procedure of obtaining a sequence of approximations
consists of three elements: of a calculational algorithm, an initial
approximation, and of additional transformations. For example, by
introducing a relaxation or damping parameter into the numerical Newton
method, one can improve the convergence of the latter [21]. Under a given
calculational algorithm, one may include control functions into an initial
approximation $P_0(x,s)$, after which all following approximations also
become dependent on $s$. This variant of introducing control functions is,
probably, the most widely used. One takes an initial Hamiltonian, or
Lagranjian, or action as depending on trial parameters that are defined as
control functions by imposing an addition condition, like the
minimal--difference condition [19,20,22-26] or the minimal--sensitivity
condition [27-32]. Such conditions are, of course, heuristic. For simple
cases, as a zero--dimensional and one--dimensional oscillators, for which
perturbative terms of arbitrary order can be obtained, one may define
control functions directly from the condition of convergence of these
terms, as $k\rightarrow\infty$ [33-35]. Finally, if a calculational
algorithm with an initial approximation have been fixed, one may introduce
control functions by subjecting the resulting asymptotic series to
additional transformations. These can be either a change of variables,
with the reexpansion of the given series in powers of new variables, or
a transformation of a series itself. An example of the former case is the
order dependent mapping, and that of the latter, the Borel--Leroy
transformation [1]. However, these transformations require the knowledge
of the analyticity properties of the sought functions itself, which is
rarely available.

To our mind, a transformation that one wishes to apply to an asymptotic
series in order to construct an analytical approach must satisfy three
main stipulations: (i) Be {\it general}, to be applicable to any function
without requiring the knowledge of its properties that are not known. The
sole assumption involved should be the existence of the sought function.
(ii) Be {\it simple}, to permit an analytical investigation. At the same
time, simplicity is usually a requisit for generality. (iii) Be {\it
invertible}, with a uniquely defined inverse transformation. This is
evidently necessary to return from a transform to the function itself.
In addition, it would be desirable to have an apparent interpretation of
the meaning of the chosen transformation.

These stipulations are satisfied by the algebraic transformation [36-38]
whose general form is
\begin{equation}
\label{4}
C_s\{ f(x)\} = a(x,s) + b(x,s)f(x) ,
\end{equation}
where $a(x,s)$ and $b(x,s)$ are any functions guarantying the uniqueness
of the inverse transformation
$$ C_s^{-1}\{ C_s\{ f(x)\}\} =
\frac{C_s\{ f(x)\} - a(x,s)}{b(x,s)} = f(x) . $$
One of the simplest variants of (4), as applied to a term $p_k(x)$ of a
sequence $\{ p_k(x)\}$, is
\begin{equation}
\label{5}
C_s\{ p_k(x)\} = P_k(x,s) = x^s p_k(x) .
\end{equation}
This variant not only simple but also it has a transparent meaning when
$p_k(x)$ is a $k$--order truncated series in powers of $x$. Then,
transformation (5) effectively increases the approximation order from $k$
to $k+s$.

Assume that, by this or that way, we have introduced control functions
constructing from an initial sequence $\{ p_k(x)\}$ a transformed sequence
$\{ P_k(x,s)\}$ with better convergence properties. Now we have to
concretize in what sense the properties of $\{ P_k(x,s)\}$ should be
better than those of $\{ p_k(x)\}$. The greatest achievement would be if
the transformed sequence $\{ P_k\}$ is such that we could notice a
relation between subsequent terms $P_k$ and $P_{k+1}$. If so, we would be
able to map the low--order terms to those of arbitrary high order. That
is, having just a few initial terms of a sequence $\{ P_k\}$, we could
extrapolate them to higher orders of $k$ defining an effective limit $P^*$
of this sequence. To formulate a relation between subsequent terms of a
sequence of approximations means to define the property of self--similarity
between these terms. This can be called the {\it approximation
self--similarity} [39-43]. To formulate the latter, we need to invoke some
further transformations. To this end, let us define an expansion function
$x(\varphi,s)$ by the equation
\begin{equation}
\label{6}
P_0(x,s) =\varphi, \qquad x=x(\varphi,s) .
\end{equation}
Then, we introduce an endomorphism
\begin{equation}
\label{7}
y_k(\varphi,s)\equiv P_k(x(\varphi,s),s) ,
\end{equation}
with an initial condition
\begin{equation}
\label{8}
y_0(\varphi,s)\equiv \varphi ,
\end{equation}
following from (6). The transformation inverse to (7) reads
\begin{equation}
\label{9}
P_k(x,s) = y_k(P_0(x,s),s) .
\end{equation}
By these definitions, the sequence $\{ y_k(\varphi,s)\}$ is bijective to
$\{ P_k(x,s)\}$. The property of self--similarity between the terms of the
sequence $\{ y_k(\varphi,s)\}$ writes [39-43] as
\begin{equation}
\label{10}
y_{k+p}(\varphi,s) = y_k(y_p(\varphi,s),s) .
\end{equation}
This is nothing but the semigroup property $y_{k+p}=y_k\cdot y_p$.
Relation (10) may remind a functional equation of renormalization group
[1,44]. However, there is here a principal difference.
Renormalization--group equations [1,44] relate a function with scaled
variables with the function itself. So, a renormalization--group equation
describes motion with respect to function variables. In our case, Eq. (10)
relates different approximations from the sequence $\{ y_k\}$. Therefore,
the approximation self--similarity (10) defines motion with respect to
approximation orders which play the role of discrete time. In the language
of dynamical theory, a dynamical system with discrete time is called a
cascade. Since the trajectory $\{ y_k(\varphi,s)\}$ of this cascade is, by
construction, bijective to the sequence of approximations $\{ P_k(x,s)\}$,
a family of endomorphisms $\{ y_k|\; k=0,1,2,\ldots\}$ can be named the
{\it approximation cascade} [45,46]. An important feature of this cascade
is that the approximation self--similarity (10) is a necessary condition
for fastest convergence [42,43].

For the purpose of developing an analytical theory, it is not convenient
to deal with discrete time. It would be desirable to pass from the
discrete index $k=0,1,2,\ldots$ to a continuous variable $t\in
[0,\infty)$. This can be done [39-42] by introducing an endomorphism
$y_t(\varphi,s)$ such that $y_t$ has the same group property,
\begin{equation}
\label{11}
y_{t+\tau}(\varphi,s) = y_t(y_\tau(\varphi,s),s) ,
\end{equation}
as $y_k$ in (10), and the values
\begin{equation}
\label{12}
y_t(\varphi,s) = y_k(\varphi,s) \qquad (t=k)
\end{equation}
at integer $t$ coincide. The so defined family of endomorphisms 
$\{ y_t|\; t\in [0,\infty)\}$ forms an {\it approximation flow}, and
conditions (11) and (12) define the {\it embedding} of a cascade into
flow [45,46]. From relation (11) with continuous time, it is easy to
derive the Lie evolution equation
\begin{equation}
\label{13}
\frac{\partial}{\partial t} y_t(\varphi,s) = v_t(y_t,s) ,
\end{equation}
with the velocity field
\begin{equation}
\label{14}
v_t(y_t,s) =\lim_{\varphi\rightarrow y_t}\lim_{t\rightarrow 0}
\frac{\partial}{\partial t} y_t(\varphi,s) .
\end{equation}
Equation (13) can be rewritten in the integral form
\begin{equation}
\label{15}
\int_{y_t}^{y_{t+\tau}}\frac{d\varphi}{v_t(\varphi,s)} = \tau .
\end{equation}
To study the properties of an approximation flow, we may invoke powerful
techniques of dynamical theory [47-50].

What we need to obtain at the end is an effective limit of the sequence
$\{ P_k(x,s)\}$. Since the latter is bijective to the trajectory 
$\{ y_k(\varphi,s)\}$ of the approximation cascade, the limit of 
$\{ P_k\}$ is in one--to--one correspondence with a stable fixed point of
the cascade [45,46]. A fixed point is defined as a zero of velocity. The
cascade velocity can be written as the Euler discretization of the flow
velocity [42,43] which reads
\begin{equation}
\label{16}
v_k(\varphi,s) = y_{k+1}(\varphi,s) - y_k(\varphi,s) +\Delta s
\frac{\partial}{\partial s} y_k(\varphi,s) ,
\end{equation}
where $\Delta s$ is a variation of a control function. Since this
variation is not known, we cannot find an exact zero of the velocity (16),
but can find only its approximate zero defining a quasifixed point. For
instance, we may put
\begin{equation}
\label{17}
\Delta s\frac{\partial}{\partial s} y_k(\varphi,s) = 0 ,
\end{equation}
which is satisfied if either $\Delta s=0$ or $\partial y_k/\partial s =0$.
In both the cases, the velocity (16) becomes
\begin{equation}
\label{18}
v_k(\varphi,s) = y_{k+1}(\varphi,s) - y_k(\varphi,s) .
\end{equation}
This and several other ways of defining quasifixed points and the related
velocities have been analysed in detail in Refs. [51-53]. The motion in the 
space of approximations, near a quasifixed point, is described by the
evolution integral (15), which can be written as
\begin{equation}
\label{19}
\int_{P_k}^{P_{k+1}^*}\frac{d\varphi}{v_k(\varphi,s)} =\tau ,
\end{equation}
where $P_k=P_k(x,s);\; P_k^*=P_k^*(x,s,\tau)$ is a quasifixed point; and
$\tau$ is a minimal time necessary to reach this quasifixed point.

Substituting the cascade velocity (18) into the evolution integral (19),
we can find a quasifixed point $P_k^*$. Then, we need to make a
transformation inverse to the algebraic transformation (4),
\begin{equation}
\label{20}
p_k^*(x,s,\tau) = C_s^{-1}\{ P_k^*(x,s,\tau)\} .
\end{equation}
The resulting approximant (20) is, as is clear from (19), a function of
$P_{k-1}$, that is of $p_{k-1}$, which can be written as
$$ p_k^* = F_k(p_{k-1}) . $$
We may repeat the renormalization procedure for $p_{k-1}$, obtaining
$$ p_k^* = F_k(F_{k-1}(p_{k-2})) , $$
and so on. After $k$ steps of such a procedure, called self--similar
bootstrap [38], we come to
$$ p_k^* = F_k\left ( F_{k-1}\left (\ldots F_1\left ( p_0\right )
\right )\ldots \right ) . $$
In short notation, the latter can be presented as a quasifixed--point
equation
\begin{equation}
\label{21}
p_k^* = F_k(p_{k-1}^*) .
\end{equation}
As far as (21) implies a $k$--step renormalization, the resulting $p_k^*$
will contain two sets
$$ \bar s_k = \{ s_1,s_2,\ldots,s_k\} , \qquad
\bar \tau_k = \{\tau_1,\tau_2,\ldots,\tau_k\} $$
of $2k$ control functions, which can be denoted as
$$ p_k^* = p_k^*(x,\bar s_k,\bar\tau_k) . $$

Now it is time to recall the main aim of the present paper, that is,
to suggest an approach for treating crossover phenomena. Therefore, we
must remember the asymptotic condition (2) and to require that the found
approximation (21) would satisfy the condition
\begin{equation}
\label{22}
p_k^*(x,\bar s_k,\bar\tau_k) \simeq f_{as}(x) \qquad 
(x\rightarrow\infty) .
\end{equation}
This defines the sets $\bar s_k=\bar s_k(x)$ and $\bar\tau_k=\bar\tau_k(x)$
of control functions. With the found control functions, we obtain the
final self--similar approximant
\begin{equation}
\label{23}
f_k^*(x) = p_k^*(x,\bar s_k(x),\bar\tau_k(x)) .
\end{equation}

What makes the present paper different from our previous publications is
the {\it systematic use of the asymptotic conditions} of type (22) for
defining control functions. The suggested procedure is designed so that to 
self--similarly sew the left and right asymptotic expansions of a function
on a given interval. For concreteness, we have considered above the sewing
procedure from the left to the right. But, as is evident, the same way can
be followed from the right to the left, that is, starting from an
asymptotic expansion at the right boundary of the interval $[0,\infty)$,
when $x\rightarrow\infty$, and then sewing the obtained approximant with
the asymptotic form at the left boundary, where $x\rightarrow 0$. In any
case, we shall come to an approximant whose structure is governed by the
quasifixed--point equation (21).

To show explicitly what is the structure of the approximant $p_k^*$, let
us take an initial expansion $p_k(x)$, as $x\rightarrow 0$, in the form of
a standard power series
\begin{equation}
\label{24}
p_k(x) =\sum_{n=0}^k a_nx^n .
\end{equation}
Employing the algebraic transform (5), we have
\begin{equation}
\label{25}
P_k(x,s) =\sum_{n=0}^k a_nx^{n+s} .
\end{equation}
Equation (6) reads
\begin{equation}
\label{26}
P_o(x,s) = a_0x^s =\varphi ,
\end{equation}
from where the expansion function is
\begin{equation}
\label{27}
x =\left ( \frac{\varphi}{a_0}\right )^{1/s} .
\end{equation}
For the endomorphism (7), we get
\begin{equation}
\label{28}
y_k(\varphi,s) = \sum_{n=0}^k a_n\left (\frac{\varphi}{a_0}\right
)^{1+n/s} .
\end{equation}
The cascade velocity (18) becomes
\begin{equation}
\label{29}
v_k(\varphi,s) = a_{k+1}\left (\frac{\varphi}{a_0}\right )^{1+(k+1)/s} .
\end{equation}
Substituting (29) into the evolution integral (19), we find a quasifixed
point $P_k^*$, after which we need to make the inverse transformation (20),
\begin{equation}
\label{30}
p_k^*(x,s,\tau) = x^{-s}P_k^*(x,s,\tau) .
\end{equation}
This results in the expression
\begin{equation}
\label{31}
p_k^*(x,s,\tau) =\left [ p_{k-1}^{-k/s}(x) -
\frac{ka_k\tau}{sa_0^{1+k/s}}x^k\right ]^{-s/k} .
\end{equation}
Let us note that when $s\rightarrow\infty$, then
\begin{equation}
\label{32}
\lim_{s\rightarrow\infty} p_k^*(x,s,\tau) = p_{k-1}(x)\exp\left (
\frac{a_k}{a_0}\tau x^k\right ) ,
\end{equation}
which explains how naturally exponentials appear in our approach [38].

The quasifixed--point equation (21), as applied to (31), gives
\begin{equation}
\label{33}
p_k^*=\left [ (p_{k-1}^*)^{1/n_k} + B_kx^k\right ]^{n_k} ,
\end{equation}
where, for brevity, the arguments of $p_k^*=p_k^*(x,s_k,\tau_k)$ are not
written down, and the notation
\begin{equation}
\label{34}
n_k\equiv -\frac{s_k}{k} , \qquad
B_k\equiv\frac{a_k\tau_k}{n_ka_0^{1-1/n_k}}
\end{equation}
is used. In the same way, we get
$$ p_{k-1}^*=\left [ (p_{k-2}^*)^{1/n_{k-1}} + B_{k-1}x^{k-1}
\right ]^{n_{k-1}} , $$
and so on down to
$$ p_2^* =\left [ (p_1^*)^{1/n_2} + B_2x^2\right ]^{n_2} $$
and
$$ p_1^* =\left ( p_0^{1/n_1} + B_1 x\right )^{n_1} , $$
where $p_0=a_0$. This show that the structure of (33) is a sequence of
nested roots. For instance, a third--order approximant looks like
$$ p_3^* =\left\{\left [\left ( p_0^{1/n_1} + B_1 x\right )^{n_1/n_2} +
B_2x^2\right ]^{n_2/n_3} + B_3x^3\right\}^{n_3} . $$
Control functions $s_k$ and $\tau_k$ are to be found from the asymptotic
condition (22). Because of relation (34), this is the same as to define
the powers $n_k$ and amplitudes $B_k$. Equalities (34) are nothing but a
change of variables, so that instead of $s_k$ and $\tau_k$ we may consider
$n_k$ and $B_k$ as new control functions. For practical purpose, we may at
once write down a $k$--order approximant in the form of (33) and to
directly define $n_k$ and $B_k$ from condition (22). If the latter gives
several solutions for control functions, then, we should opt for that
solution which leads to the decrease of $B_k$ with increasing $k$. This
follows from (33), from where it is evident that $p_k^*$ tends to a fixed
point $p^*$ if and only if $B_k\rightarrow 0$, as $k\rightarrow\infty$.

Before passing to the consideration of complex physical problems using the
method we have described, let us illustrate it by a simple example of the
one--dimensional quartic anharmonic oscillator. Consider the dimensionless
ground--state energy $e(g)$ as a function of the coupling, or anharmonicity, 
parameter $g\in[0,\infty)$. In the weak--coupling limit, when
$g\rightarrow 0$, perturbation theory results [54] in the expansion
\begin{equation}
\label{35}
e(g)\simeq  a_0 + a_1 g + a_2 g^2 + a_3 g^3 + a_4 g^ 4 ,
\end{equation}
where
$$ a_0=\frac{1}{2} , \qquad a_1 = \frac{3}{4}, \qquad a_2=-\frac{21}{8},
\qquad a_3 = \frac{333}{16} , \qquad a_4=-\frac{30885}{128} . $$
In the strong--coupling limit, when $g\rightarrow\infty$, the asymptotic
behavior is known [55,56] to be
\begin{equation}
\label{36}
e(g) \simeq A_0 g^{1/3} + A_1g^{-1/3} + A_2 g^{-1} + A_3g^{-5/3} ,
\end{equation}
with the coefficients
$$ A_0=0.667986, \quad A_1=0.143669, \quad A_2=-0.008628, \quad
A_3=0.000818 . $$
Starting from the linear approximation $p_1(g)=a_0 + a_1(g)$ from (35), we
find
$$ p_1^*(g,n_1,B_1) =\left ( a_0^{1/n_1} + B_1g\right )^{n_1} , $$
which is the first approximation from (33). Requiring the validity of the
asymptotic condition
$$ p_1^*(g,n_1,B_1) \simeq A_0 g^{1/3} \qquad (g\rightarrow\infty) , $$
we find $n_1=\frac{1}{3},\; B_1=A_0^3=0.298059$. From here, we could
recalculate $s_k$ and $\tau_k$ using relations (34), however, this is not
necessary, since, as is explained above, now $n_k$ and $B_k$ play the role
of control functions, and what we need finally are exactly $n_k$ and
$B_k$. The quantity $n_k$ can be called a {\it crossover index} and
$B_k$, {\it crossover amplitude}. With the found $n_1$ and $B_1$, we
define, analogously to (23), the first--order self--similar approximant
$$ e_1^*(g) \equiv p_1^*\left ( g,\frac{1}{3}, A_0^3\right )  $$
for the ground--state energy, which writes
\begin{equation}
\label{37}
e_1^*(g) =\left ( a_0^3 + A_0^3 g\right )^{1/3} .
\end{equation}
Comparing the values of (37) with numerical results [55] that can be
treated as exact, we see that the maximal error, for $g\geq 0$, of Eq.
(37) is $-6.8\%$ occurring at $g\approx 0.7$.

Following the same way, we find the second--order self--similar crossover
approximant
\begin{equation}
\label{38}
e_2^*(g) = \left [\left ( a_0^{9/2} + Cg\right )^{4/3} + 
B_2g^2\right ]^{1/6} ,
\end{equation}
where two first terms in the asymptotic expansion (36) are used, and
$C=0.1971,\; B_2=A_0^6=0.0888$. The maximal error of (38) is $-2.9\%$ at
$g\approx 0.3$. Continuing the procedure, we get the third--order
approximant
\begin{equation}
\label{39}
e_3^*(g) = \left\{\left [\left ( a_0^{81/14} + C_1g\right )^{4/3} + 
C_2 g^2\right ]^{7/6} + B_3g^3\right\}^{1/9} ,
\end{equation}
with $C_1=0.1116,\; C_2=0.0784,\; B_3=A_0^9=0.02648$. The maximal error of
(39) is $-1.7\%$ at $g\approx 0.1$. In the fourth order, we obtain
\begin{equation}
\label{40}
e_4^*(g) =\left\{\left\{\left [\left ( a_0^{243/35} + D_1 g\right )^{4/3}
+ D_2g^2\right ]^{7/6} + D_3 g^3\right\}^{10/9} + B_4 g^4\right\}^{1/12} ,
\end{equation}
where $D_1=0.0625,\; D_2=0.05005,\; D_3=0.0131$, and $B_4=A_0^{12}=0.00789$. 
The maximal error of (40) is $-1.3\%$ at $g\approx 0.1$. The sequence is
uniformly convergent, which can be seen from the monotonic decrease of
errors from about $7\%$ to $1\%$.  

We would like to emphasize that our aim in the present paper is to suggest
a {\it systematic analytical} method permitting one to derive explicit
expressions describing crossover phenomena. The advantage of having accurate
analytical formulas, as compared to numerical results of numerical methods, 
is in the simplicity of analyzing such formulas with respect to the
variation of parameters entering these formulas. Also, having an analytical 
formula corresponding to a measurable quantity often gives more information 
about the studied system than just numbers. As an example, we may mention
the geometric spectral inversion in quantum mechanics [57,58].

\section{Fr\"ohlich Polaron}

The Fr\"ohlich optical polaron problem [3] is an interesting physical
example of a crossover about which there existed a controversy lasting for
around 30 years. Some researchers, analysing the polaron ground--state
energy $e(\alpha )$ as a function of the electron-phonon coupling parameter
$\alpha$, found an indication to a phase transition from a state of freely
moving weak--coupling polaron to a localized state of a strong--coupling
polaron (see discussion in [15]). One of the first such indications has
been suggested by Gross [59]. However, as modern investigations show [4],
there is no phase transition in the polaron problem but the latter is an
example of a classical crossover between the weak--coupling and
strong--coupling limits.

In the weak-coupling limit, the ground state of polaron has an asymptotic
behaviour
\begin{equation}
\label{41}
e(\alpha)\simeq a_1\alpha +a_2\alpha ^2+a_3\alpha ^3,\qquad 
\left( \alpha \rightarrow 0 \right) , 
\end{equation}
with three well--established terms [15,60,61], in which
$$
a_1=-1,\qquad a_2=-1.591962 \times 10^{-2},\qquad 
a_3=-0.806070\times 10^{-3}. 
$$
In the strong--coupling limit, Miyake [62,63] obtained 
\begin{equation}
\label{42}
e(\alpha)\simeq A_0\alpha^2 + A_2 + A_4\alpha^{-2} \qquad 
\left( \alpha \rightarrow \infty \right) , 
\end{equation}
where
$$
A_0=-0.108513,\qquad A_2=-2.836,\qquad A_4=-4.864. 
$$
The terms of the weak--coupling expansion are known here with a better
precision than those of the strong--coupling expansion. In addition, the
coefficients, $a_k$ decrease as $k$ increases, while $A_k$ increase with
$k$. Therefore, here we have to construct self--similar approximations
from the right to the left, that is, starting from the perturbative
expression (42), we find a self--similar approximant $e_k^*(\alpha)$, with
control functions defined from the asymptotic condition
$$ 
e_k^*(\alpha) \simeq e_{as}(\alpha) \qquad (\alpha\rightarrow 0) , 
$$
in which $e_{as}(\alpha)$ is given by (41). The accuracy of the found
self--similar approximants $e_k^*(\alpha)$ can be evaluated by comparing
them with the values $e(\alpha)$ obtained by Monte Carlo numerical
calculations [4,64]. As usual, the accuracy of self--similar crossover
approximants is the worst in an intermediate region, where a
weak--coupling expansion is sewed with a strong--coupling one. For 
the polaron energy $e_k^*(\alpha)$, the maximal error occurs at
$\alpha\approx 10$.

The first--order crossover approximation gives
\begin{equation}
\label{43}
e_1^*(\alpha) =  -\alpha\left ( 1 + B\alpha^2\right )^{1/2} ,
\end{equation}
with $B=A_0^2=0.011775$. The maximal error of (43) is $-10.5\%$. The
second--order approximant is
\begin{equation}
\label{44}
e_2^{*}(\alpha)= -\alpha \left[ 1 + \alpha\left ( B_0 + B_2\alpha^2
\right )^{3/2} \right ]^{1/4}, 
\end{equation}
where $B_0=0.159468,\; B_2=A_0^{8/3}=0.002679$. The maximal error of (44)
is $-4.54\%$. Finally, we may find the third--order self--similar
approximant taking account of all three known terms in (41). This yields
\begin{equation}
\label{45}
e_3^*(\alpha)= -\alpha\left\{ 1 + \alpha\left[ C_0 + \alpha\left (
C_1 + C_2\alpha^2\right )^{3/2} \right ]^{5/4}\right \}^{1/6} , 
\end{equation}
where $C_0=0.152804,\; C_1=0.049617$, and $C_2=A_0^{16/5}=0.000819$. The
maximal error of (45) is $-1.5\%$. Again, we see that, with increasing
approximation order, the accuracy of the found crossover approximants
improves, from an error of about $10\%$ to that of about $1\%$. The very
simple formula (45) gives the same accuracy as the Feynman variational
calculations [3,65].

\section{Kondo Effect}

One of the most remarkable examples of crossover phenomena is given by 
the Kondo effect [2]. The behavior of the system, consisting of a local
impurity spin and conduction electrons, interacting by means of an 
antiferromagnetic exchange of strength $J$, changes from asymptotically
free at high temperatures to that of the impurity screened by electronic
lump, via the crossover region whose onset is characterized by the Kondo
temperature estimated as $T_k=D\exp (-1/2J)$, where $D$ stands for the
Fermi--energy of electrons. We consider below only the case of a
single--channel Kondo model with the impurity local moment equal to $1/2$.
Most of our knowledge about the problem comes from the exact Bethe ansatz
solution [66,67], from the field--theoretical renormalization group (RG)
[68-70], and from the Wilson numerical renormalization group [71]. It was
pointed out in Ref.[72], that the Bethe--ansatz solution cannot be
extrapolated beyond the coupling constant $J$ of order one (see also
[73]). Field--theoretical RG results are valid only at $J<1$ as well. On
the other hand, the strong--coupling limit, $J\rightarrow\infty$, of the
Kondo model was considered in Ref.[72]. Only the numerical RG treatment of
the Kondo  problem is valid, formally, for arbitrary $J$. We suggest below
a simple analytical approach valid for arbitrary $J$.

Within the framework of the field--theoretical RG in its application to
the Kondo crossover, the central role is played by the so-called invariant
charge or effective electron-electron coupling $J_{inv}$ [68-70], measuring
the intensity of electron--electron interactions via the impurity spin. The
field--theoretical Gell--Mann--Low $\beta$--function, could be defined
using the perturbation theory in the weak--coupling limit [68-70],
\begin{equation}
\label{46}
\beta (J)\simeq -2J^2 + 2J^3, \qquad (J\ll 1), 
\end{equation}
or by means of a sophisticated bosonization technique in the
strong--coupling limit [72]: 
\begin{equation}
\label{47}
\beta (J)\simeq -c,\qquad c\approx 0.377,\qquad 
(J\rightarrow \infty). 
\end{equation}
The left crossover approximation, satisfying by design both known limits, 
can be obtained giving an improved, self--similarly renormalized 
Gell--Mann--Low function: 
\begin{equation}
\label{48}
\beta^*(J)=-2J^2 \left( 1+ \frac{\tau}{2}J \right )^{-2} ,
\qquad \tau\equiv \sqrt{\frac{8}{c}}=4.607 , 
\end{equation}
and $J_{inv}$ is given [68-70] by the equation
$$
\int_J^{J_{inv}}\frac{dg}{\beta^*(g)} = \ln \left( 
\frac{\omega}{D} \right ) .
$$
The last integral can be calculated explicitly and the result may be 
presented in the form
$$
\Phi (J_{inv})=\ln \left( \frac{\omega}{T_k}\right) , 
$$
\begin{equation}
\label{49}
\Phi(z) = \frac{1}{2z} - \frac{\tau}{2}\ln(z) - \frac{\tau^2}{8}\ z, 
\end{equation}
where $\omega$ stands for the typical external parameter of the problem
(temperature, magnetic field) and $T_k$ is the typical internal energy
scale, or the Kondo temperature 
\begin{equation}
\label{50}
T_k = D J^{\tau/2} \exp\left ( - \frac{1}{2J} + \frac{\tau^2}{8} J
\right ) .
\end{equation}
This expression for the Kondo temperature has the same form as the famous
Wilson numerical RG result, 
$$
T_k = \widetilde{D}(J) (2J)^{1/2}\exp \left [ -
\frac{1}{2J} + 1.5824(2J)\right ] , 
$$
where $\widetilde{D}(J)$ is known to have a power series expansion in 
$J$ [71]. The origin of the linear correction in the exponential, can be
traced, therefore, to the strong--coupling limit. To our knowledge, other
analytical approaches, including the Bethe ansatz solution, cannot capture
it. The effective interaction determined by Eq. (49), grows to infinity as 
$\omega$ goes to zero in agreement with the numerical RG and strong--coupling
limit [71,72,74].
   
\section{One--Dimensional Antiferromagnet}

An extreme caution is needed when any kind of perturbative or
non--perturbative approach is applied to the one--dimensional Heisenberg
antiferromagnet of arbitrary spin $S$. Even such a general method as Bethe
ansatz fails for $S>1/2$. Nevertheless, the crossover approach can be of
use in this situation.

\subsection{Autocorrelation Function.}

The Bethe ansatz, despite its failure in the general case of $S>1/2$, 
allows one to find the magnetic properties of the Heisenberg
antiferromagnetic (AF) spin chains of arbitrary spin, when a maximum of
two deviations is allowed from the completely aligned (ferromagnetic)
state [75]. The magnetization curve and pair correlations had been
obtained explicitly for a strong magnetic field, close to the spin--flop
transition. The expression for the autocorrelation function 
$\Theta_0=\left\langle S_0^zS_0^z\right\rangle $, as a function of the
number $N$ of spins $S$, has a very simple and transparent form. In Ref.
[75], an equivalent quantity
$$
F_0(N) = \frac{\Theta _0}{S^2}-\left ( \frac{S_T^z}{NS}\right )^2
$$
is presented (and compared to numerical data) as a function of a parameter 
$\sigma$ (demagnetization),
$$
\sigma \equiv 1 - \frac{S_T^z}{NS}\; . 
$$
Here $S_T^z$ stands for the magnetization of a spin--flop phase and is
controlled by the magnetic field $h$, so that close to the saturation field 
$h_s=4S$ [75] one has: 
$$
\frac{S_T^z}{NS} = 1 - \frac{2}{\pi S}\left ( 1 - \frac{h}{h_s} 
\right )^{1/2}. 
$$
Finally, $F_0$ is presented in the following form:
\begin{equation}
\label{51}
F_0(\sigma)\simeq \frac{\sigma}{S} - \sigma^2 + \frac{1}{8}
\pi^2 S^2 (2S - 1)^2\sigma^4\qquad (\sigma \ll 1)\; . 
\end{equation}
The last term in (51), proportional to $(2S-1)^2$, clearly distinguishes
an extra--contribution from the so--called $C$--states, typical for 
$S\geq 1$ and absent for $S=1/2$, with high probability of having two spin
deviations on the same site [75]. An exact, independent on $N$, value 
$$
F_0 = \frac{2}{3}\; , \qquad \left ( \sigma = 1,\; S =1 \right ) , 
$$
is known too [75], and can be used as an asymptotic condition. Let us
continue the expansion (51) from the region of small $\sigma$ to the
region of $\sigma\sim 1$, along the stable trajectory, ending at
$\sigma=1$ at the value $F_0=\frac 23$. In order to extend the validity of
(51) for $S\geq 1$ , let us add to (51) one more trial term $\sim-\sigma^6$, 
and find the corresponding effective time $\tau$ from the crossover
condition at the boundary point. The self--similar bootstrap procedure
leads to the following crossover approximations : 
\begin{equation}
\label{52}
F_0^{*}(\sigma) = \frac{\sigma}{S(1+\sigma S)},\qquad 
S = \frac{1}{2}\; , 
\end{equation}
\begin{equation}
\label{53}
F_0^*(\sigma) = \frac{\sigma}{S}\exp \left [ - \sigma S\exp \left ( 
- \frac{1}{8}\pi^2 S^2 (2S - 1)^2\sigma^2 \exp \left ( -
\frac{A\sigma^2}{S^2(2S-1)^2} \right ) \right ) \right] ,\quad S\geq 1, 
\end{equation}
where $A=8\tau/\pi^2=0.253$. At $S=1$, $F_0^*(\sigma)$ agrees,
both qualitatively and quantitatively, with the data of Fig. 3 from
Ref. [75], the maximal error being $\approx 4\%$. The behavior of
$F_0^*(\sigma)$ for $S=1/2$ and $S=1$ is qualitatively similar (universal
regime) only as $\sigma\rightarrow 0\; (h\rightarrow h_s)$, where 
$C$--states are suppressed by a magnetic field, and is different for all
finite $\sigma$ ($h<h_s$) due to the contribution from $C$--states
(non--universal regime). An onset of the regime dominated by $C$--states
may be related to the inflection point of the curve $F_0^*(\sigma)$ for 
$\sigma\sim 0.5$, emerging for $S=1$, and absent for $S=1/2$.

Consider the case of $h\rightarrow h_s,\; S=1$. Upon rapid (instant)
switching off magnetic field down to the value $h=0$, the correlations
between spins should change from the behavior typical of the universal
regime to that of the non-universal regime, dominated by $C$--states. One
may suspect, therefore, that the typical time interval of relaxation of
physical properties, such as the pair correlation function of spins and
the magnetization, would be radically different for $S=1/2$ and $S=1$. To
be more specific, consider another physical property $\Gamma$, an effective
interaction of two spin flips, which can be presented in the vicinity of the
saturation field as a function of $S$ and $h_s-h$ [76]: 
\begin{equation}
\label{54}
\Gamma (S,h)\sim 
\frac{2}{\sqrt{2}(\pi S\delta)^{-1} + 1 - 1/S}\; ,\qquad
\delta = \sqrt{4-\frac{h}{S}}. 
\end{equation}
As $h\rightarrow h_s$, $\Gamma$ remains positive for arbitrary spins and
the system enters the universal regime, when all equilibrium physical
properties for arbitrary spins could be derived from the "Bose--gas with
repulsion" model [76-78]. As $h$ is instantly set to zero, the magnetization 
should readjust itself from the values near to the saturation to the zero
magnetization. The expression (54) for $\Gamma$ can still be used in this
situation as an estimate of the effective interaction of two spin flips,
at least at the initial stage of relaxation. Then, from (54), it follows
that $\Gamma (S=1/2,h=0)\approx -3.64$, i.e. acquires the negative
sign different from that of $\Gamma (S=1,h=0)\approx 8.89$. The function 
$\Gamma (S,h=0)$ has a peak at $S=1$, then saturates as $S\rightarrow\infty$
to the positive value $2$. Negative sign of $\Gamma(S=1/2)$ means an
attraction of two spin flips and rapid collapse of the magnetized state
to the state without magnetization, while positive $\Gamma(S=1)$ means
repulsion and much longer relaxation time for the magnetization. I.e., we 
can expect an {\it anomalously slow relaxation} of the magnetization for
spin $1,$ after instant switching off magnetic field from the value
close to saturation down to zero, compared to the case of spin $1/2$.

\subsection{Ground--State Energy}

Spin--wave theory gives for the ground state energy $E$ of the Heisenberg
AF in the one--dimensional case the expansion in powers of inverse spin
$1/S$ (see [79] and Refs. therein): 
\begin{equation}
\label{55}
E \simeq -S^2\left( 1+\frac{\gamma}{2S} \right ) ,\qquad 
\gamma \approx 0.7. 
\end{equation}
The self--similarly renormalized expression, following from (55), is 
\begin{equation}
\label{56}
E^* = -S^2 \exp\left (\frac{\gamma}{2S}\tau \right ), 
\end{equation}
and at $\tau =1$, $E^{*}(S=1)=-1.419$, approximating the "exact" numerical
result $-1.401$ [80] with an accuracy of $1.285\%$, which is an improvement
as compared to the error $-3.64\%$ of expression (55), corresponding to
"bare" spin waves. For $S=2$, $E^{*}=-4.765$, in excellent agreement with
the "exact" numerical result $-4.761$ [81].

The error, calculated for the renormalized expression (56) for $S=1/2$, is
equal to $13.54\%$, becoming much worse than $-4.06\%$ for the bare spin
waves, as compared to the exact value $E_0=-0.44315$ [30]. An attempt to
improve the result for $S=1$, choosing the effective time $\tau$ from the
exact result at $S=1/2$, gives the error of $-5\%$ as $S=1$, suggesting,
that between $S=1/2$ and $S=1$, some new physical mechanism enters the play,
invalidating our attempt to match smoothly the ground state energies for
quantum spins based only on renormalization of spin--wave formula. On the
other hand, a successful estimation of the ground state energy for $S=1,2$,
based on $1/S$--expansion, suggests that a similar mechanism works for all
$S\geq 1$ and quite an accurate estimate can be obtained from formula (56).

Motivated by the existence of exact results for the autocorrelation
function, assume that the ground state energy could be expanded around
$E_0$ in powers of $\left( S-\frac{1}{2}\right)$, i.e introduce a trial 
$\left( S-\frac{1}{2}\right)$--expansion around the exact solution at
$S=\frac{1}{2}$:
\begin{equation}
\label{57}
E\sim -\left[ \left| E_0\right| +A \left( S-\frac 12\right) \right]\; ,
\qquad S\rightarrow \frac {1}{2}\; . 
\end{equation}
The coefficient $A$ will be determined by matching Eq. (57) with the
expression for the ground state energy $E$, as $S\rightarrow\infty$:
\begin{equation}
\label{58}
E\sim -S^2 \qquad \left( S\rightarrow \infty \right) . 
\end{equation}
Following the standard prescriptions of Section II, we obtain the left
crossover approximation 
\begin{equation}
\label{59}
E^* = - \left [ \sqrt{\left |E_0\right|} + \left( S-\frac{1}{2}
\right) \right]^2,\qquad  A=2\left| E_0\right|^{-1/2}, 
\end{equation}
with $E^*(S=1)=-1.359$, approximating the exact result with the
percentage
error of $-3\%$, being only slightly better than the spin--wave result. In
order to check the idea about similar mechanisms, forming the ground state
energy for $S\geq 1$, rewrite (59) in the form 
\begin{equation}
\label{60}
E^* = - \left[ S^2 + \left( 2\sqrt{\left| E_0\right| }-1\right) S + \left(
\left| E_0\right| - \sqrt{\left| E_0\right| }+1/4\right) \right] , 
\end{equation}
and consider (60) as another form of $1/S$--expansion. Applying to Eq. (60)
the procedure of self--similar renormalization, we obtain: 
\begin{equation}
\label{61}
E^{**} = - S^2\exp \left( \frac{2\sqrt{\left| E_0\right| }-1)}{S}\right) , 
\end{equation}
with $E^{**}(S=1)=-1.393$, and the error $-0.57\%$. For $S=2$, 
$E^{**}=-4.72$, and the error is equal to $-0.86\%$. We again conclude, that 
for the ground state energy a simple crossover formula exists, covering
the region from large spins to the small quantum spin $S=1$.

\subsection{Haldane Gap}

Haldane [5] conjectured the existence of radically different elementary
excitation spectra for arbitrary integer and half--odd--integer $1d$
Heisenberg spins, the former case being gapped with the smallest value 
of the gap $\Delta$ at  $k=\pi$, while the latter case is gapless. In the
limit of large $S$, Haldane used an approximate mapping onto the $O(3)$
non--linear sigma--model, leading to the following behavior of the gap [5]: 
\begin{equation}
\label{62}
\Delta \sim S^2\exp (-\pi S), \qquad S\rightarrow \infty . 
\end{equation}
Strictly speaking, formula (62) describes only the "slow" part of the 
full dependence and does not take into account the "fast" part, describing
the gap oscillations with changing spin, with zeros at half--odd--integer
spins and maxima at integer values. Nowadays, it is established beyond the
reasonable doubt [82], that for the half--odd--integer spins 
\begin{equation}
\label{63}
\Delta \equiv 0, \qquad  \left ( S= \frac{1}{2},\; \frac{3}{2},\;
\frac{5}{2}\ldots \right ) . 
\end{equation}
For small integer spins $S=1,2$, the values of the gap are known from an
extensive numerical calculations. We suggest below a simple way to estimate 
$\Delta$ for arbitrary integer spins, based on the self--similar
renormalization of a trial $\left( S-\frac 12\right)$--expansion for the
Haldane gap and on the knowledge of the asymptotic form (62), as
$S\rightarrow\infty$, together with the demand for the absence of the gap
for half--odd--integer spins.

Let us write the trial expansion for the gap in the vicinity of the point 
$S=\frac 12$ in the following form, satisfying the condition 
$\Delta\left( S=\frac 12\right) =0$ : 
\begin{equation}
\label{64}
\Delta \sim a_2\left( S-\frac 12\right)^2 + a_3\left( S-\frac 12\right)^3
+ a_4 \left( S-\frac 12\right)^4 + a_5 \left( S-\frac 12\right)^5 +
\ldots \qquad \left( S\rightarrow \frac 12\right) , 
\end{equation}
where $a_k$ are positive. Following the general prescriptions of Section II, 
one can self--similarly renormalize (64) to the form
\begin{equation}
\label{65}
\Delta^* = a_2\left ( S- \frac{1}{2}\right )^2 \left [ 1 -
C_1\left ( S -\frac{1}{2}\right ) \right ]^{n_1}\exp \left \{ C_2\left (
S -\frac{1}{2}\right )^2\left [ 1 - C_3\left ( S - \frac{1}{2}\right )
\right ]^{n_2}\right \} \; .
\end{equation}
We require that (65) agrees with (62), as $S\rightarrow\infty$. Demand also, 
that at $S=3/2$, $\Delta=0$. Choosing the unknown coefficients and powers
in (65) so that to satisfy the required conditions, we may come to
\begin{equation}
\label{66}
\Delta^* = \left( S-\frac{1}{2}\right )^2 \exp \left [ - 
\frac{(S - \frac{1}{2})^2}{S-\frac{3}{2}} \right ] ,
\end{equation}
where the value at $S=\frac{3}{2}$ is defined as the limit from the right,
$S\rightarrow\frac{3}{2}+0$. At $S=1$, $\Delta^*=0.412$, in good agreement
with the "exact" numerical value $0.4105$ [80]; at $S=2$, $\Delta^*=0.025$, 
agreeing by the order of magnitude with numerical value $0.085(5)$ [81]
or, even in better agreement with the value $0.05,$ quoted in Ref. [83].
Formula (66) can be generalized requiring that the exponential in Eq.
(66) should have the form of an expansion in $1/S$ as $S\rightarrow\infty$
and zeros at $S=\frac{5}{2},\;\frac{7}{2},\;\frac{9}{2},\ldots$, which
yields 
\begin{equation}
\label{67}
\Delta^* = \left( S-\frac 12\right) ^2\exp \left[ -\left( \ \frac{
\left( S-\frac 12\right) ^2}{\left( S-\frac 32\right) }+\frac{\left( S-\frac
32\right) ^3}{\left( S-\frac 52\right) ^3}+\frac{\left( S-\frac 52\right) ^4 
}{\left( S-\frac 72\right) ^5}+\frac{\left( S-\frac 72\right) ^5}{\left(
S-\frac 92\right) ^7}+ \ldots \right) \right] ,
\end{equation}
where again the values at $S=(2n+1)/2$ are defined as the limits from the
right. The value of the gap at $S=1$, given by Eq. (67), remains practically 
the same as above, while at $S=2,\; \Delta^*=0.068$. The gap, when described 
by formula (67), practically vanishes for all integer $S\geq 3$, in agreement
with the conclusion of Ref. [81].

\subsection{Other Characteristics}

Self--similar approximants can be constructed for other characteristics as
well. Here we briefly mention only a couple of examples. Staggered
magnetization $\Sigma$ of the antiferromagnetic anisotropic
Ising--Heisenberg model of spin--$1/2$, as a function of the anisotropy
parameter $\gamma$ (equal to zero for the Ising and one for the Heisenberg
model), can be presented as the expansion, valid at small $\gamma$ [84,85]: 
\begin{equation}
\label{68}
\Sigma (\gamma) \simeq 1 - \gamma^2 - \frac{1}{4}\gamma^4 - \ldots\; . 
\end{equation}
As $\gamma =1$, according to Ref. [85], the long--range order parameter
$\omega_\infty =\Sigma^2(\gamma)$ should disappear, i.e. $\Sigma(1)=0$.
Let us continue expression (68), from the region of $\gamma \ll 1$, to the
region of $\gamma \sim 1$, satisfying the boundary condition for the
disappearance of the long--range order. Then the left crossover approximation 
\begin{equation}
\label{69}
\Sigma^*(\gamma) = \left ( 1- \gamma^2 \right ) \exp \left( -
\frac{1}{4}\gamma^4 \right ) 
\end{equation}
satisfies the right boundary condition. Comparing $\omega _\infty^* =
(\Sigma^*(\gamma))^2$, with the extrapolation of numerical data, presented
in Fig.30 of Ref. [85], we found that they almost coincide.

At zero temperature, the dispersion, known from the linear spin--wave
theory, is modified by the factor $Z$ in the spin--wave velocity, and the
expansion for  $Z$ in powers of the inverse coordination number $1/z$ was
obtained [86,87]: 
$$
Z\simeq 1 + \frac{1}{4Sz} + \frac{3}{16Sz^2} + \ldots \;, 
$$
We continue this expression from the small values of $1/z$ to arbitrary
$z$, while the value of spin is fixed,  and determine the effective time
$\tau$ from the exact value of $Z=\pi/2$, at $S=1/2,\; z=2$ [87].
The left crossover approximation has the following form: 
\begin{equation}
\label{70}
Z^* = 1 + \frac{1}{4Sz}\exp \left( \frac{3}{4z}\tau \right) , \qquad 
\tau = 2.21. 
\end{equation}
At $S=1/2$, $z=4$ (square lattice) we obtain from Eq. (70) $Z^{*}=1.189$,
in excellent agreement with the results obtained by different methods [87]. 
At $S=1/2$, $z=6$, the case corresponding to a simple cubic lattice,
$Z=1.11$, i.e the quantum corrections to the spin velocity remain important.

\section{Collective Excitations}

The knowledge of the elementary excitation spectrum is one of the key
points for the description of many--body problems. Dealing with this
extremely complicated problem, one often encounters the situation when 
the elementary excitation spectrum $\omega(k)$ is known for two different
regions of wave vector $k$. In the hydrodynamic region, $k\rightarrow 0$, 
the form $\omega(k)$ could be determined either from experiment or
theoretically. In the short--wave region, $k\rightarrow\infty$, a dispersion 
corresponding to free particles should recover. Using the self--similar
renormalization, it is possible to reconstruct $\omega(k)$ for arbitrary
$k$. Consider some problems of this kind, frequently occurring in condensed 
matter physics.

\subsection{Bogolubov Spectrum}

The case of a linear in $k$ spectrum, as $k\rightarrow 0$, and of
quasi--free massive particles, as $k\rightarrow\infty$, is of the 
most general type when Bose systems are considered. This kind of behavior
is inherent to Bose systems and does not depend on the details of an
interaction potential [6]. Consider the case of an anomalous sound
dispersion, corresponding to an instability of the spectrum, as $k$
increases [7]. The following asymptotic expressions are available: 
$$
\omega(k) \simeq ck\left ( 1+\gamma k^2 \right ) , \quad
\gamma >0 \qquad  \left ( k\rightarrow 0 \right ) , 
$$
\begin{equation}
\label{71}
\omega(k) \simeq \frac{k^2}{2m^*} \qquad
\left ( k\rightarrow\infty \right ) . 
\end{equation}
Here $c$ is the velocity of sound, $\gamma$ is responsible for the
instability of the spectrum and $m^*$ is an effective mass. The left
crossover approximation can be derived following the standard prescriptions
of Section II, which gives the result identical to the Bogolubov spectrum
of a weakly non--ideal Bose gas: 
\begin{equation}
\label{72}
\omega^*(k)=ck\sqrt{ 1+ \left( \frac{k}{2m^*c}\right )^2}\; . 
\end{equation}
Note, that in distinction from the microscopic Bogolubov approach, valid for
a diluted Bose system, formula (72) may be used for arbitrary densities,
assuming that the parameters $c$ and $m^*$ are taken from experiment. It
looks rather intriguing that the same formula (72), that is usually derived 
with some lengthy calculations, can be immediately obtained by 
self--similarly interpolating the simple asymptotic expressions (71).

\subsection{Liquid Helium Spectrum}

Consider the case when more terms in the hydrodynamic limit are available,
but the "anomalous" dispersion coefficient $\gamma$ is very small. Also,
free particles are replaced by quasi--free "dressed" particles with an
effective mass $m^*$. The asymptotic behaviour of the spectrum is as
follows: 
$$
\omega(k) \simeq ck \left( 1 + \gamma k^2 - \delta k^4 \right ) ,\quad
\gamma \approx 0,\quad \delta >0 \qquad \left ( k\rightarrow 0\right ) , 
$$
\begin{equation}
\label{73}
\omega (k) = \frac{k^2}{2m^*} \qquad
\left ( k\gg  1 \AA^{-1} \right ) . 
\end{equation}
This situation is typical for liquid $He^4$, where 
$\gamma=0\pm 0.05\AA^2,\; \delta =0.29\pm 0.03\AA^4$ [88], and
$m^*=2-3m(He^4)$ [89]. 
The crossover approximant derived from (73) reads
\begin{equation}
\label{74}
\omega^*(k) = ck \left[ \left ( \exp\left ( -\delta k^4 \right )
\right )^6 + \left ( \frac{k}{2m^*c} \right )^6 \right ]^{1/6} .
\end{equation}
Expression (74) generalizes the Bogolubov spectrum (72). The main difference 
originates from the region of the intermediate $k\sim 1$. Formula (74)
describes the experimental data for the elementary excitation spectrum of
liquid $He^4$ [88] both qualitatively, predicting the existence of roton
minimum even for the "bare" mass $m^*=m(He^4)$, and quantitatively, with
the maximal percentage error of about $20\%$. The value of the effective
mass $m^*=2-3m(He^4)$ may have some relation to the formation of
two--particle and three--particle correlated states [90]. We took above
for estimates the value of the sound velocity equal to $2.4\cdot 10^4cm/s$.

Our approach to deriving the spectrum for arbitrary $k$ corresponds to the
Feynman approach [65], when only the information about the short--wave and
long--wave parts of the structure factor $S(k)$ are used. Then, instead of
the phenomenological Feynman formula $\omega(k)=k^2/2mS(k)$, we apply the
self--similar renormalization. The result is a Bogolubov--type formula.
Thus, a bridge between the Bogolubov and Feynman approaches to the spectrum 
of Bose systems [91] is established. Formula (74) is better qualitatively
than the original Bogolubov spectrum (71), since it predicts the
maxon--roton region, and better quantitatively than Feynman formulae,
especially in the roton region. Here, the Feynman formula works with
an error of about $100\%$, while Eq. (74), in the worst case, gives an
error about $10\%$. Our formula (74), is a three--parametric representation 
of the spectrum of liquid $He^4$, with parameters $c,\;\delta$ and $m^*$
coming from the regions of long, intermediate and short--wave lengths,
respectively.

The case of a stable sound--like spectrum, as $k\rightarrow 0$, and of
quasi--free particles, as $k\rightarrow\infty$, can also correspond to a
collective--excitation branch in liquid $He^3$ [92]. The following
asymptotic expressions are available: 
$$
\omega(k) \simeq ck \left ( 1 - \left |\gamma\right | k^2 \right ) , 
\qquad \left( k\rightarrow 0\right) , 
$$
\begin{equation}
\label{75}
\omega(k) \simeq \frac{k^2}{2m^*}\; ,
\qquad  \left ( k\rightarrow\infty \right ) . 
\end{equation}
By analogy with the case of $He^4,$ we find the  spectrum 
\begin{equation}
\label{76}
\omega^*(k) = ck \left [ \left ( \exp (- \left | \gamma \right | k^2)
\right )^4 + \left ( \frac{k}{2m^*c} \right )^4 \right ]^{1/4}. 
\end{equation}
However, since phonons in liquid $He^3$ are intertwined, in the region of
intermediate wave vectors, with other collective excitations, it is
impossible to observe rotons. 

\subsection{Spectrum with Gap}

Assume that the spectrum has a gap, as $k\rightarrow 0$, and possesses a
minimum at this point, while, as $k\rightarrow\infty$, it becomes linear: 
$$
\omega(k) \simeq \Delta + \alpha k^2,\quad \alpha >0 \qquad
\left ( k\rightarrow 0\right ) , 
$$
\begin{equation}
\label{77}
\omega(k) \simeq vk, \qquad
\left ( k\rightarrow\infty \right ) . 
\end{equation}
The left crossover approximation can be readily obtained, leading to the
expression 
\begin{equation}
\label{78}
\omega^*(k) = \Delta \sqrt{1+\left ( \frac{vk}{\Delta} \right )^2}, 
\end{equation}
analogous to the spectrum of the Bardeen--Cooper--Schrieffer model of
superconductivity.

\subsection{Dynamical Scaling}

The characteristic frequency $\omega_c(\zeta,k)$, appearing in the
dynamical scaling hypothesis [8,93] and proportional to an inverse
characteristic relaxation time of an order parameter, has two
asymptotic
forms, depending on the ratio $k/\zeta$, where $\zeta$ stands for the
inverse correlation length. We shall discuss below only the behavior of
density--density correlations in liquid systems [8]. Asymptotic expansions
in the hydrodynamic regime $(k/\zeta\ll 1)$  and in the fluctuation regime
$(k/\zeta\gg 1)$ are known:
$$
\omega_c (\zeta,k) \simeq D_Tk^2 \left [ 1 + B \left ( \frac{k}{\zeta}
\right )^2 + \ldots \right ] ,\quad B>0 , \qquad  
\left ( \frac{k}{\zeta} \ll 1 \right ) , 
$$
\begin{equation}
\label{79}
\omega_c(\zeta,k) \simeq Ak^z \left [ 1 + A' \left ( 
\frac{\zeta}{k}\right )^2 + \ldots \right ] ,\qquad 
\left ( \frac k\zeta \gg 1\right ) , 
\end{equation}
where $D_T$ is a thermal diffusivity and $z$ is the dynamical
critical index, which cannot be determined self--consistently within the
framework of the dynamical scaling. The value of $B$ is estimated as 
$B=1$ [94], or $B=3/5$ [95]. 

Assume that the values of $z$ and $A$ are known. Then one can reconstruct
the analytical expression for the characteristic frequency for arbitrary
$k/\zeta$ obtaining the following left crossover approximation
\begin{equation}
\label{80}
\omega_c^*(\zeta,k) = D_T k^2 \left [
1 + C\left (\frac{k}{\zeta}\right )^2\right ]^n , 
\end{equation}
where
$$
C = \zeta^2\left ( \frac{A}{D_T}\right )^{2/(z-2)} , \qquad 
n = \frac{z}{2} - 1 .
$$
For $z=3$ [8], we obtain $n=1/2$. If now we plug into expression (80)
the dependencies of $D_T\sim\epsilon^{\gamma-\alpha}$ and
$\zeta\sim\epsilon^\nu$ on the distance $\epsilon$ from the critical 
point [8], then we recover immediately the well--known relation between 
the critical indices $z,\;\gamma,\;\nu$ and $\alpha$, that is,
$z=2+(\gamma-\alpha)/\nu$ (all definitions are standard and may be found in 
Ref. [8]), which represents one of the central results of the dynamical
scaling hypothesis. From this scaling relation, the dynamical critical
index could be estimated from the values of three other indices. 

\section{Conclusion}

We suggested a general approach to describing crossover phenomena of
arbitrary nature. The approach permits one to construct an accurate
approximation for a function in the whole domain of its variable from
asymptotic expansions near the boundaries. The minimal information needed
for obtaining a self--similar interpolation formula is two terms of an
expansion near one of the boundaries and the limiting value at another
boundary. Having only three such terms, it is already possible to get a
reasonable approximation for the sought function in the total crossover
region. When more terms are available, the procedure may be continued
improving the accuracy of approximations. An important feature of the
method is that the self--similar crossover approximants always preserve
the correct structure of the asymptotic expansions at both boundaries of
the interpolation region. This is a clear advantage of the self--similar
approach as compared to often used heuristic interpolations that may spoil
the structure of the asymptotic expansions.

The possibility of obtaining accurate approximations from an extremely
scarce information, when no other methods work, is based on the following
three points: (i) The idea of self--similar renormalization group treating
the transfer from one approximation to another as the evolution of a
dynamical system, approximation cascade. (ii) The requirement that
this evolution be invariant with respect to algebraic transformations. 
(iii) The use of control functions providing the stability and 
convergence of procedure.

Control functions introduced under the algebraic self--similar
renormalization play, for the crossover problem, the role of effective
crossover indices and effective crossover times. Depending on whether we
start the renormalization procedure from an expansion either near the left
or near the right boundary, we may distinguish the left and right crossover
indices and, respectively, the left and right crossover times. Similarly,
the resulting expressions for the sought function may be called the left and
the right crossover approximations.

The form of the resulting self--similar approximations depends on the
properties of the asymptotic expansions used. Mathematically equivalent
expansions lead to the same form of crossover approximations. For example,
compare the ground state energy of the Fr\"ohlich polaron as a function 
of the coupling parameter and the spectrum of collective excitations as 
a function of the wave vector. The weak coupling series in powers of the
coupling parameter is analogous to the long--wave spectrum in powers of 
the wave vector. The strong coupling limit for the optic polaron is similar 
to the short--wave limit for the collective spectrum. As a result of this,
the crossover approximation for the polaron energy has the same dependence
on the coupling parameter as the crossover approximation for the collective 
spectrum on the wave vector. Thus, physically different quantities may have 
the same mathematical representation as functions of the corresponding
variables. Keeping this in mind, we may say that there exist the 
{\it classes of universality} of crossover phenomena.

It is worth emphasizing that the crossover approximations derived by
applying the approach developed usually combine good accuracy with
simplicity. This suggests that the self--similar renormalization provides
a natural tool for extracting the maximal information from very short
perturbative series that are impossible to analyze by other methods.
Moreover, this makes us to think that self--similarity, in some sense, is 
hidden in asymptotic series. This is why the self--similar renormalization
becomes a natural effective tool of extracting such a hidden information.
Different physical examples presented in this paper prove as well that
this is also a general tool applicable to arbitrary crossover phenomena.

\vspace{5mm}

{\bf Acknowledgement}

\vspace{1mm}

We are grateful to E.P. Yukalova for many discussions and advice.
Financial support from the National Science and Technology Development
Council of Brazil and from the University of Western Ontario, Canada, is
appreciated.

\end{document}